\documentclass[journal]{IEEEtran}
\usepackage{amsmath,amssymb,amsfonts}
\usepackage{graphicx}
\usepackage{xurl}
\usepackage{booktabs}
\usepackage{array}
\usepackage{textcomp}
\usepackage{placeins}
\providecommand{\tightlist}{\setlength{\itemsep}{0pt}\setlength{\parskip}{0pt}}
\begin{document}

\title{Physical-Layer Fingerprint-Space Capacity Analysis\\
for 100BASE-TX Devices\\
in IIoT}

\author{Chenming~Zhang and Aiqun~Hu,~\IEEEmembership{Senior Member,~IEEE}%
\thanks{Chenming Zhang is with Duke Kunshan University, Kunshan 215316, China, and also with the School of Cyber Science and Engineering, Southeast University, Nanjing 210096, China (e-mail: cz229@duke.edu).}%
\thanks{Aiqun Hu is with the State Key Laboratory of Mobile Communication, Southeast University, Nanjing 210096, China (e-mail: aqhu@seu.edu.cn). (Corresponding author: Aiqun Hu.)}%
\thanks{This work has been submitted to the IEEE for possible publication. Copyright may be transferred without notice, after which this version may no longer be accessible.}%
}

\markboth{Preprint}%
{Zhang and Hu: Physical-Layer Fingerprint-Space Capacity Analysis}

\IEEEtitleabstractindextext{%
\begin{abstract}
Industrial Internet of Things (IIoT) networks widely adopt Ethernet technologies, such as 100BASE-TX, for industrial communications. As industrial networks continue to scale, reliable device authentication becomes increasingly important for preventing device impersonation and unauthorized access. Physical-layer fingerprinting (PLF) exploits device-dependent fingerprint features in transmitted signals and provides a hardware-based approach for terminal authentication. However, the distinguishable space supported by 100BASE-TX physical-layer fingerprints and its capacity boundary remain largely unexplored. To analyze the capacity of physical-layer fingerprints, this paper proposes a nonlinear and impulse-response model (NAIM) that characterizes device-dependent waveform differences in 100BASE-TX transmitted waveforms. The nonlinear component captures steady-state level deviations, while the impulse-response component describes the transition response during level transitions. The 100BASE-TX transmitter waveform requirements, the observation resolution determined by noise and analog-to-digital conversion (ADC) quantization, and the target bit-error ratio (BER) constrain the admissible fingerprint space. Under the NAIM model, the fingerprint-space capacity of 100BASE-TX terminals is derived as approximately $2.96\times10^{10}$ distinguishable states. Experiments on signals collected from 48 NICs under two cable conditions estimate a Gaussian-equivalent empirical capacity from the measured inter-device and within-device variations. Under the 5-m cable condition, empirical capacity and closed-set identification consistently rank the three NIC models, and a larger empirical capacity yields higher identification accuracy. These results demonstrate that the proposed capacity analysis provides a pre-deployment assessment for physical-layer fingerprinting in IIoT.
\end{abstract}

}

\maketitle
\IEEEdisplaynontitleabstractindextext

\section{Introduction}\label{i.-introduction}

\IEEEPARstart{A}{t the} field-access layer of the Industrial Internet of Things (IIoT), sensors, controllers, and edge terminals commonly connect to industrial networks through 100BASE-TX [1], [2]. Its mature protocol ecosystem and relatively low deployment cost have sustained widespread use in existing production lines and industrial edge devices.

However, as the number of IIoT nodes grows and industrial networks become increasingly interconnected, reliable device authentication becomes essential for protecting network access. Attackers can impersonate legitimate devices and gain network access through physical ports, thereby bypassing upper-layer authentication and logical isolation mechanisms to steal sensitive data, tamper with control commands, or disrupt industrial operations [2]. This type of wired intrusion poses a serious threat to the secure and reliable operation of IIoT systems.

Existing IIoT access control systems primarily rely on digital credentials such as MAC addresses and digital certificates. These credentials authenticate software-presented identities, but cannot reliably verify that a credential comes from the claimed physical device. If a credential is cloned or leaked, an access-control system may therefore accept an impersonating device [2]. This risk is especially relevant to resource-constrained and widely distributed industrial terminals that are difficult to update. These limitations highlight the need for authentication mechanisms that are directly associated with the physical characteristics of connected devices.

In recent years, device fingerprinting has emerged as a promising approach to device authentication by exploiting device-dependent variations embedded in transmitted signals. Because these variations are tied to the physical device, device fingerprinting provides a hardware-based complement to conventional software credentials.

Gerdes et al.~provided an early physical-layer identification framework for wired Ethernet [3]; subsequent works have further extracted device-related information from adaptive filtering responses, spectra, access waveforms, and correlation spectra [4], [5], [6], [7]. More recent studies on 1000BASE-T have enhanced wired fingerprint authentication from the perspectives of interpretable feature modeling and open-set rejection, respectively [8], [9]. Together, these studies establish that wired Ethernet signals contain measurable device-dependent variations for physical-layer authentication. As industrial networks continue to scale, an important question remains: whether the available fingerprint space is sufficiently large to accommodate the increasing number of connected devices.

Existing wired fingerprint studies primarily characterize identification performance over fixed device sets. Such results characterize whether the measured devices can be separated, whereas fingerprint-space capacity concerns the scale of waveform states that remain distinguishable under a specified observation condition. Quantifying this capacity requires characterizing how device-dependent waveform variations are distributed within the physically allowable region and how small these variations can be resolved under practical observation conditions.

An explicit waveform model is therefore required to parameterize device-dependent variation within this space. The MLT-3 signal of 100BASE-TX uses positive, zero, and negative levels. In actual network interface cards (NICs), the steady-state levels deviate from their ideal values, while each transition is shaped by the analog transmit path. NAIM represents these two forms of variation through a steady-state nonlinear mapping and an impulse-response representation, respectively.

The 100BASE-TX waveform requirements bound admissible parameter variation, while the observation resolution sets the minimum distinguishable change. Because simultaneous parameter variations can interact through the common output waveform, fingerprint-space capacity is determined by the jointly feasible parameter combinations rather than by independent per-parameter ranges.

Measurements from real NICs are further used to characterize practical device separability in the same fingerprint coordinate representation. Inter-device spread relative to repeated-measurement variation is used to estimate Gaussian-equivalent empirical capacity, while cross-day closed-set identification evaluates device-level separability.

Accordingly, this paper develops a model-based framework for analyzing the capacity of 100BASE-TX physical-layer fingerprints. The main contributions are as follows:

\begin{enumerate}
\def\labelenumi{\arabic{enumi})}
\item
  We develop a model-based capacity analysis framework for 100BASE-TX physical-layer fingerprints, where NAIM is introduced to represent device-dependent waveform variations through steady-state level and impulse-response components.
\item
  We formulate fingerprint-space capacity by jointly considering the feasible parameter ranges imposed by the selected 100BASE-TX waveform requirements, the resolution of the observation model, and the joint feasibility of multi-parameter waveform variations.
\item
  We evaluate the framework on 48 NICs under two cable conditions and characterize measured fingerprint variations through Gaussian-equivalent empirical capacity and multi-day closed-set identification.
\end{enumerate}

Under the specified waveform and observation settings, the fingerprint space contains approximately \(2.96 \times 10^{10}\) distinguishable states. Measurements from 48 NICs under two cable conditions further report device separability through empirical capacity, and evaluate the relationship between identification performance and empirical capacity through multi-day closed-set identification. These results indicate that the proposed capacity analysis provides a pre-deployment assessment for applying physical-layer fingerprinting in IIoT.

\FloatBarrier
\section{Related Work}\label{ii.-related-work}

Device fingerprinting distinguishes network terminals through hardware-dependent variations in their transmitted signals. Existing wired Ethernet studies have primarily investigated how these variations are represented as device fingerprints and how the resulting fingerprints are used for authentication. Related research has also examined authentication limits and capacity in physical-layer identification.

\subsection{Wired Device-Fingerprint Extraction and Modeling}\label{a.-wired-device-fingerprint-extraction-and-modeling}

Gerdes et al. provided an early physical-layer identification framework for wired Ethernet [3]. By processing analog transmit signals, they showed that measurable differences among NICs can support physical-layer device identification. This result established the transmitted waveform as a source of wired device fingerprints.

Subsequent studies have extracted fingerprint features such as spectra and correlation spectra from transmitted Ethernet signals. For example, J. Liu et al.~extracted spectral fingerprints from 100BASE-TX signals [5]; Y. Liu et al.~constructed device features using the correlation spectrum of 1000BASE-T signals [7]; and Suski et al.~extracted fingerprints from signals transmitted at the Ethernet physical layer [6]. Together, these methods derive device-related fingerprint features from measured Ethernet signals for device identification.

A different representation of device-related variation can be obtained from the response of the transmission process. Li et al. used adaptive filtering to estimate a device-related response from Fast Ethernet signals [4]. This approach relates the fingerprint to a response of the transmission process rather than using only spectral or waveform statistics.

Recent wired studies have further expanded how Ethernet physical-layer fingerprints are modeled and constructed. Zhong et al. proposed an interpretable high-pass-filter model for 1000BASE-T and related its filter parameters to waveform shape [8]. Y. Hu et al. extracted 100-Mbps Ethernet fingerprints using variational mode decomposition and the Hilbert--Huang transform [10], and later used Ethernet physical-layer signals to construct the ES-PUF [11].

\subsection{Authentication Evaluation and Fingerprint-Space Capacity}\label{b.-authentication-evaluation-and-fingerprint-space-capacity}

Most wired studies evaluate identification in a closed set, where every test device belongs to the registered set [8]. Open-set recognition additionally requires the system to reject unregistered devices [9], [12]. Wireless radio-frequency fingerprinting (RFF) research has also treated receiver and channel variation as a distinct robustness problem [13], [14], [15]. Together, these studies evaluate identification and rejection under different device sets and observation conditions.

From an information-theoretic perspective, Gungor and Koksal studied fundamental limits of RF-fingerprint-based authentication [16], while Wang et al. introduced user capacity to characterize the number of users that wireless physical-layer identification can accommodate under specified identification requirements [17]. Wang et al. also developed a modeling and validation framework for wireless physical-layer identification [18]. Together, these studies extend physical-layer identification research from experimental identification performance to theoretical limits and user-scale analysis.

For 100BASE-TX, however, existing fingerprinting studies have primarily focused on fingerprint extraction and identification, while quantitative analysis of the distinguishable fingerprint-space size has received limited attention.

\FloatBarrier
\section{100BASE-TX Waveform Formation and the Nonlinear and Impulse-Response Model}\label{iii.-100base-tx-waveform-formation-and-the-nonlinear-and-impulse-response-model}

To analyze how large a distinguishable fingerprint space 100BASE-TX can support, this section first traces the device-dependent waveform variations that persist after standard digital encoding. These variations appear in two forms: deviations among the steady-state MLT-3 levels and differences in the waveform following each level transition. These two waveform characteristics reflect different aspects of the analog transmit path. NAIM models them separately so that the resulting parameters retain direct interpretations in terms of steady-state levels and transition dynamics.

\subsection{100BASE-TX Transmit Chain and Waveform Differences}\label{a.-100base-tx-transmit-chain-and-waveform-differences}

As illustrated in Fig. 1, the 100BASE-TX transmission link can be simplified into two stages: standard digital encoding and a device-dependent analog path. In the digital domain, 4B/5B encoding, scrambling, NRZI encoding, and MLT-3 conversion produce a three-level symbol sequence in \{+1, 0, \ensuremath{-}1\}. The resulting three-level sequence then enters the device-dependent analog path, which comprises the line driver, output shaping, and magnetic coupling and determines the transmitted waveform at the active output interface (AOI), where the waveform requirements considered in this work are specified for a 100 \ensuremath{\Omega} differential load [19]--[21].

\begin{figure}[!t]
\centering
\includegraphics[width=\columnwidth]{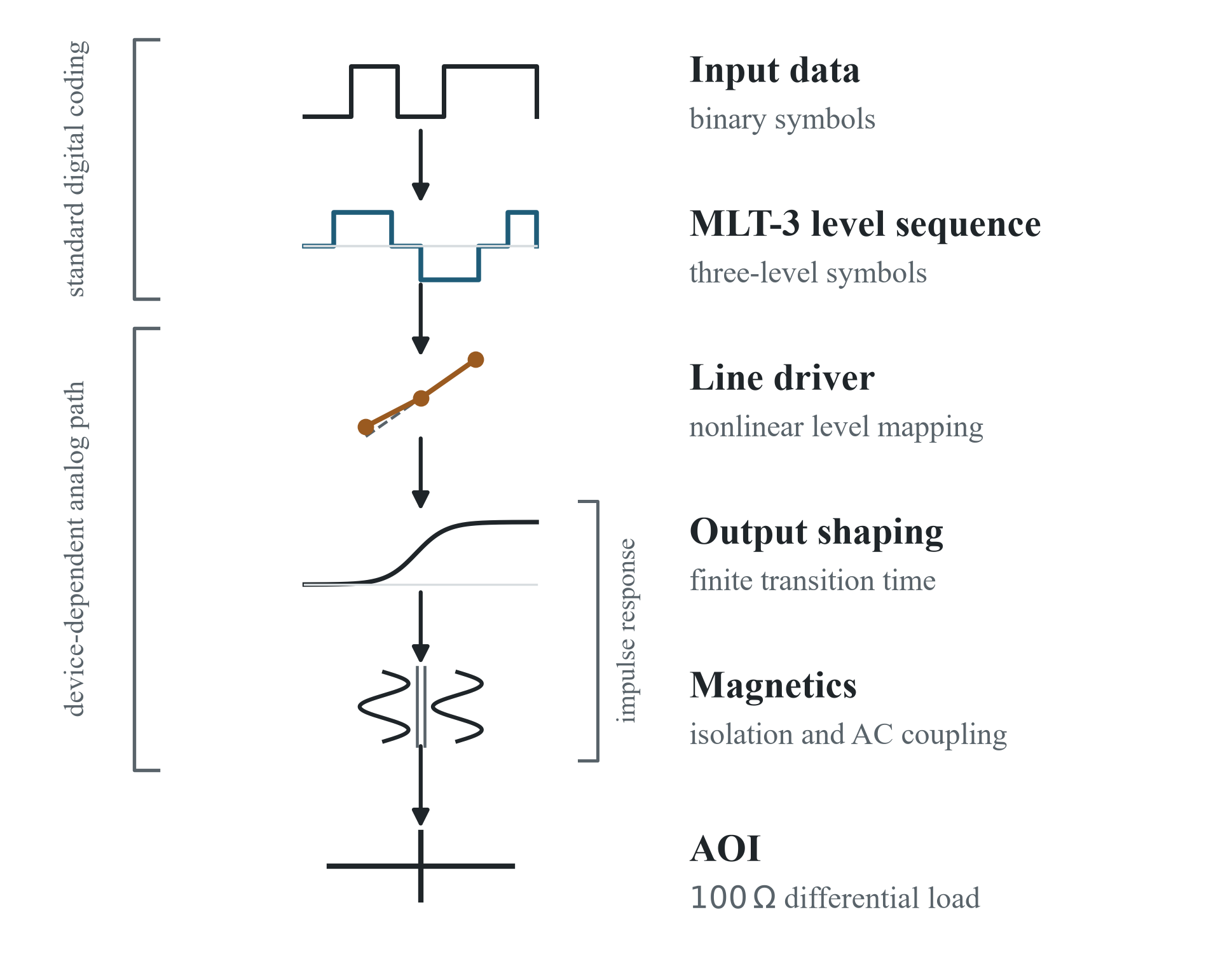}
\caption{Simplified 100BASE-TX transmit chain and waveform formation. Standard digital encoding yields an MLT-3 symbol sequence; the device-dependent analog path then produces the transmitted waveform at the active output interface (AOI) under a 100 \ensuremath{\Omega} differential load.}
\end{figure}

The 100BASE-TX physical-layer standard specifies the 4B/5B, scrambling, NRZI, and MLT-3 operations in the digital domain; under compliant implementations, the same input follows the same three-level symbol rules [19], [21], [20]. Device-dependent waveform differences therefore arise primarily from the subsequent analog path. Variations in the analog path affect both the steady-state output levels and the transition dynamics: line-driver behavior is reflected in the level amplitudes, while output shaping and magnetic coupling contribute to finite rise/fall behavior and post-transition settling. Even for the same MLT-3 symbol sequence, these variations produce different transmitted waveforms across devices. At the waveform level, the differences appear in two forms: steady-state level deviations change the amplitudes and symmetry of the MLT-3 plateaus, whereas transition-response variation changes the rise, decay, and settling following each transition.

Fig. 2 shows these two waveform characteristics for the 48 NICs measured on the first day under the 0.5-m condition: the steady-state levels in Fig. 2(a) and the estimated finite impulse response (FIR) in Fig. 2(b).

\begin{figure*}[!t]
\centering
\includegraphics[width=\textwidth]{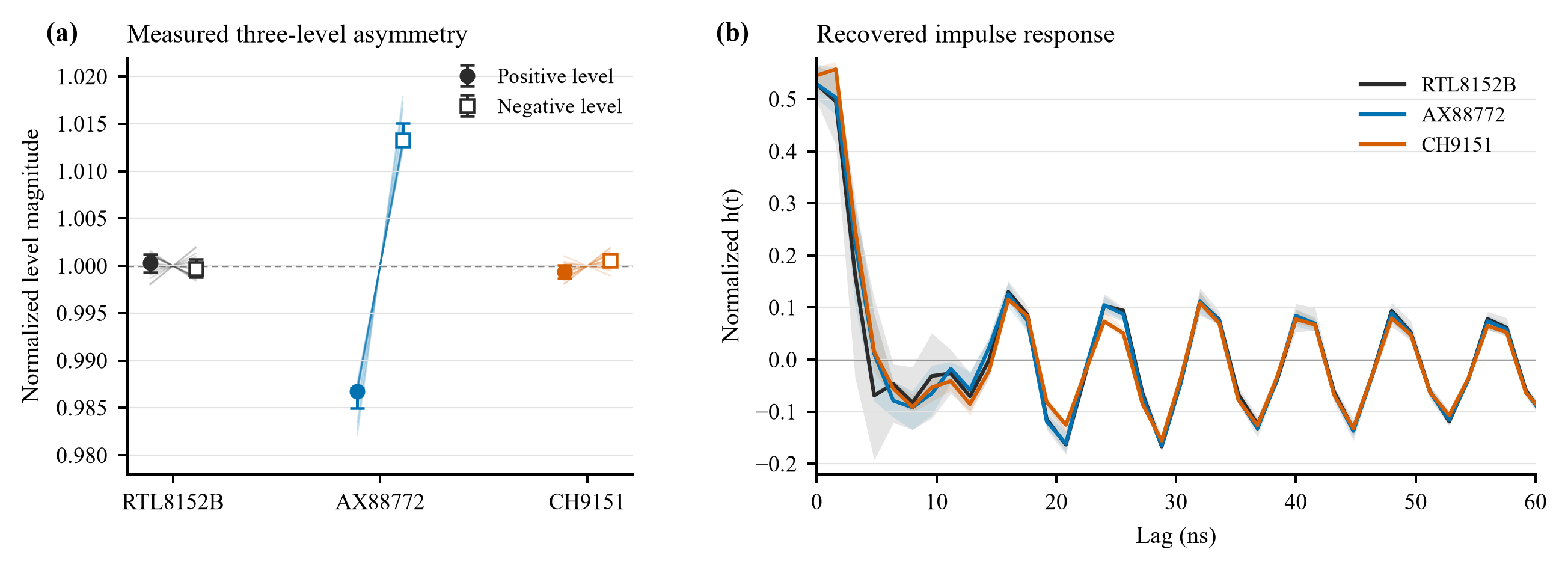}
\caption{Measured level amplitudes and estimated FIR response. (a) Positive and negative level amplitudes. (b) Estimated FIR response.}
\end{figure*}

As shown in Fig. 2(a), the positive and negative plateaus are not perfectly symmetric, with the deviation being particularly visible for AX88772. Fig. 2(b) also shows differences among the estimated FIR responses, particularly in their post-peak settling behavior. These observations motivate the two components of NAIM: a nonlinear level mapping for the steady-state amplitudes and an impulse-response model for the transition dynamics.

Because the steady-state plateau value depends on the current MLT-3 level, it is represented by the memoryless mapping \(\Gamma( \cdot )\). In contrast, the finite rise/fall behavior and subsequent settling depend on preceding transitions and are represented by the impulse response \(h[k]\). Together, these two components form a Hammerstein-type NAIM [22]: \(\Gamma( \cdot )\) determines the device-dependent level sequence, while \(h[k]\) describes the transition dynamics. Let \(x[k]\) be the ideal MLT-3 level sequence and \(n[k]\) the observation noise; the noiseless model output and observed waveform are then

\begin{equation}
y_{0}[k] = h[k]\ast\Gamma\bigl(x[k]\bigr),\qquad y_{\mathrm{obs}}[k] = y_{0}[k] + n[k].
\label{eq:naim}
\end{equation}

where \(\ast\) denotes linear convolution and \(n[k]\) is the observation noise.

\subsection{MLT-3 Steady-State Level Mapping}\label{b.-mlt-3-steady-state-level-mapping}

The plateau asymmetry observed in Fig. 2(a) cannot in general be represented by a single linear gain, because the positive and negative levels need not be symmetric about the center. Taking the center plateau as the reference, we therefore represent the steady-state level dependence by the polynomial mapping

\begin{equation}
\Gamma(x) = \sum_{p = 1}^{P}a_{p}x^{p},
\label{eq:gamma-poly}
\end{equation}

and require that \(\Gamma(0)=0\). Here, \(x\) is the ideal MLT-3 input relative to the center level, taking values in \(\{+1, 0, -1\}\); \(a_{p}\) is the coefficient of the \(p\)-th-order term; and \(P\) is the truncation order of the polynomial. The constraint \(\Gamma(0)=0\) indicates that, with the center plateau as the reference, a zero input corresponds to a zero relative output. The three input levels then correspond to

\begin{equation}
\Gamma(+1)=\mu_{+}-\mu_{0},\qquad \Gamma(-1)=\mu_{-}-\mu_{0}
\label{eq:gamma-levels}
\end{equation}

where \(\mu_{+}\), \(\mu_{0}\), and \(\mu_{-}\) denote the steady-state levels of the positive, center, and negative plateaus, respectively.

The MLT-3 alphabet is \(x\  \in \ \left\{ - 1,\ 0,\  + 1 \right\}\). On these three points, any odd power satisfies \(x^{2r + 1} = x\), and any positive even power satisfies \(x^{2r} = x^{2}\). Therefore, all odd-order terms and even-order terms in \eqref{eq:gamma-poly} can be merged separately as

\begin{equation}
\Gamma(x) = A_{\mathrm{odd}}\, x + A_{\mathrm{even}}\, x^{2}.
\label{eq:gamma-two}
\end{equation}

For example, when written up to the fifth order, \(A_{\mathrm{odd}} = a_{1} + a_{3} + a_{5}\) and \(A_{\mathrm{even}} = a_{2} + a_{4}\); on the three-level alphabet, higher-order expansions still merge into these two coefficients. This order reduction is entirely determined by the three-level alphabet and requires no additional assumption about the polynomial order of the device circuit.

From \eqref{eq:gamma-levels} and \eqref{eq:gamma-two}, it directly follows that

\begin{equation}
\begin{aligned}
A_{\mathrm{odd}} &= \frac{\mu_{+} - \mu_{-}}{2} \\
A_{\mathrm{even}} &= \frac{\mu_{+} + \mu_{-}}{2} - \mu_{0}
\end{aligned}
\label{eq:aodd-aeven}
\end{equation}

The three-level alphabet therefore reduces the steady-state mapping to two effective parameters with direct waveform interpretations. \(A_{\mathrm{odd}}\) is the differential amplitude (half-difference) of the positive and negative plateaus; under a three-level input, it absorbs both the linear gain and the odd-order terms that cannot be separated further. \(A_{\mathrm{even}}\) is the midpoint offset of the positive and negative levels relative to the center, and reflects even-symmetry imbalance. These two parameters capture the steady-state level variation, while the time-dependent waveform changes following each transition are represented by the impulse-response component developed next.

\subsection{Impulse Response and Waveform Parameter Vector}\label{c.-impulse-response-and-waveform-parameter-vector}

After the steady-state level is set by \(\Gamma(\cdot)\), the remaining post-transition dynamics are represented by the impulse response \(h[k]\). Over the 1--200 MHz analysis band, h[k] is expanded using a fixed family of damped-sinusoidal candidates that represent transition components at different oscillatory time scales and decay rates. For the m-th candidate, \(f_{m}\), \(Q_{m}\), and \(\varphi_{m}\) denote the center frequency, quality factor, and quadrature phase, respectively; their numerical settings are summarized in Table II. The candidate transient waveform is defined as

\begin{equation}
\psi_{m}[k]=\exp\Bigl(-\frac{\pi f_{m}}{Q_{m}f_{s}}k\Bigr)\sin\Bigl[\frac{2\pi f_{m}}{f_{s}}k+\varphi_{m}\Bigr]
\label{eq:basis-pulse}
\end{equation}

where \(f_{s}\) is the sampling frequency.

Each candidate waveform \(\psi_{m}[k]\) is mean-removed, restricted to the 1--200 MHz analysis band, and normalized to obtain the fixed basis waveform \(\overline{\psi_{m}}[k]\). The resulting family forms a band-pass decomposition because each damped sinusoid is concentrated near its center frequency. A nonredundant subset of these basis waveforms is shared across all devices; device dependence therefore enters through the expansion coefficients \(b_{m}\) while the basis waveforms remain fixed.

With \(h_{0}[k]\) denoting the fixed baseline response shared by all devices, the modeled impulse response is

\begin{equation}
h[k]=h_{0}[k]+\sum_{m=1}^{M}b_{m}\overline{\psi}_{m}[k]
\label{eq:impulse}
\end{equation}

Here, \(b_{m}\) determines the magnitude and sign of the m-th basis contribution. The fixed baseline \(h_{0}[k]\) provides the common unit-gain component of the response, while the zero-mean basis waveforms represent device-dependent changes in transition shape. The center frequency determines the oscillatory time scale, \(Q_{m}\) controls the decay rate, and the two quadrature phases provide complementary transient components. Thus, the center frequency, quality factor, and phase define the fixed basis waveforms, whereas the coefficients \(b_{m}\) carry the device-dependent impulse-response variation.

For each recorded waveform, an FIR response is estimated by ridge regression and projected onto the fixed basis to obtain the coefficients \(b_{m}\). Because the waveform is observed after the NIC, cable, terminal load, and measurement chain, the estimated response reflects their joint effect under the corresponding observation condition.

Collecting the steady-state mapping parameters and the impulse-response expansion coefficients, we obtain

\begin{equation}
\boldsymbol{\theta}=\bigl[A_{\mathrm{odd}},\,A_{\mathrm{even}},\,b_{1},\ldots,b_{M}\bigr]^{\mathrm{T}}
\label{eq:theta}
\end{equation}

Equation~\eqref{eq:theta} collects the parameters of the memoryless level mapping and the impulse-response model into a finite-dimensional NAIM parameterization. The device-dependent parameter vector \(\boldsymbol{\theta}\) contains the two steady-state parameters and the impulse-response coefficients, while the fixed baseline \(h_{0}[k]\) is excluded from \(\boldsymbol{\theta}\). This construction keeps the two waveform components distinct while representing them in a common parameter space for the fingerprint-space analysis developed next.

\FloatBarrier
\section{Fingerprint-Space Capacity Estimation}\label{iv.-fingerprint-space-capacity-estimation}

The NAIM parameters are estimated from the recorded waveforms and then varied to generate output waveforms. The waveform requirements limit the parameter ranges, while noise and ADC resolution limit the differences that can be observed. Because several parameters affect the same output waveform, only combinations that satisfy the requirements together are counted. The resulting number of distinguishable combinations defines the fingerprint-space capacity.

\subsection{Fingerprint Coordinate Construction}\label{a.-fingerprint-coordinate-construction}

The parameter vector in \eqref{eq:theta} contains the two steady-state level parameters and the band-pass (BP) coefficients. These two groups do not vary independently: part of the BP-coefficient change is a linear companion of the level-parameter change. Treating the two groups as independent coordinates would therefore count that shared variation twice.

The two groups are standardized separately on the training data, giving the level matrix \(\mathbf{Z}_{\mathrm{NL}} = \bigl[ Z_{A_{\mathrm{odd}}},\, Z_{A_{\mathrm{even}}} \bigr]\) and the BP matrix \(\mathbf{Z}_{\mathrm{BP}}\). Linear regression of the BP coordinates on the level coordinates removes the BP variation already explained by the levels:

\begin{equation}
\mathbf{Z}_{\mathrm{BP}} = \mathbf{1}\boldsymbol{\beta}_{0}^{\mathrm{T}} + \mathbf{Z}_{\mathrm{NL}}\mathbf{B}_{\mathrm{reg}} + \mathbf{E},
\label{eq:bp-reg}
\end{equation}

where \(\mathbf{1}\) is a column of ones, \(\boldsymbol{\beta}_{0}\) is the intercept, \(\mathbf{B}_{\mathrm{reg}}\) is the regression coefficient matrix, and \(\mathbf{E}\) is the residual matrix. The residual matrix is then standardized with the training-set mean and standard deviation. Let \(Z_{\mathrm{tr}}(\cdot)\) denote this transformation:

\begin{equation}
\mathbf{Z}_{\mathrm{BP,res}} = Z_{\mathrm{tr}}(\mathbf{E}).
\label{eq:bpres}
\end{equation}

The rows of \(\mathbf{Z}_{\mathrm{BP,res}}\) are the residual impulse-response coordinates; \(\mathbf{z}_{\mathrm{BP,res}}\) denotes one such row. Combining these coordinates with the standardized level coordinates yields the joint coordinate \(\mathbf{q}\):

\begin{equation}
\mathbf{q} = \bigl[ z_{A_{\mathrm{odd}}},\, z_{A_{\mathrm{even}}},\, \mathbf{z}_{\mathrm{BP,res}}^{\mathrm{T}} \bigr]^{\mathrm{T}}.
\label{eq:q}
\end{equation}

This coordinate is used next to evaluate directional resolution and feasible ranges.

\subsection{Active Output Interface and Waveform Requirements}\label{b.-active-output-interface-and-waveform-requirements}

To determine the admissible range of \(\mathbf{q}\), the NAIM output waveform is constrained by selected 100BASE-TX transmitter requirements defined at the active output interface (AOI). IEEE 802.3 Clause 25 incorporates ANSI INCITS 263-1995 by reference and specifies the corresponding modifications [19], [23]; the selected requirements considered here assume a 100-\ensuremath{\Omega} differential load. The experimental waveforms are recorded at the observation point after the cable and measurement chain and are used to estimate the NAIM parameters. The standard-derived AOI requirements are then applied to the corresponding NAIM output waveform in the fingerprint-space calculation.

Evaluating these waveform requirements requires mapping the analysis coordinate \(\mathbf{q}\) back to the NAIM waveform-parameter vector \(\boldsymbol{\theta}\). Inverting the standardization and regression relations yields the raw model-parameter vector

\begin{equation}
\boldsymbol{\theta}_{\mathrm{raw}} = \bigl[ A_{\mathrm{odd,raw}},\, A_{\mathrm{even,raw}},\, \mathbf{b}^{\mathrm{T}} \bigr]^{\mathrm{T}}
\label{eq:theta-raw}
\end{equation}

The two steady-state components are then converted to volts using the reference amplitude \(A_{0}\) defined in Section V. Let \(V_{0}=1\,\mathrm{V}\) and \(g=V_{0}/A_{0}\). The corresponding parameter mapping is

\begin{equation}
\boldsymbol{\theta}=\mathrm{diag}(g,g,1,\ldots,1)\,\boldsymbol{\theta}_{\mathrm{raw}}.
\label{eq:theta-volt}
\end{equation}

Only the two steady-state components are scaled by \(g\); the BP coefficients remain unchanged. Substituting \(\boldsymbol{\theta}\) into \eqref{eq:naim} yields the noiseless waveform used to evaluate the selected AOI requirements.

The primary waveform requirements are constructed from four groups of standard-derived time-domain requirements [19], [23]:

\begin{enumerate}
\def\labelenumi{\arabic{enumi})}
\tightlist
\item
  The differential peak output is 0.950--1.050 V;
\item
  the overshoot does not exceed 5\%, and any overshoot or undershoot transient decays to within 1\% of the steady-state value within 8 ns after the transition begins;
\item
  the ratio of the positive and negative amplitudes is 0.98--1.02;
\item
  the rise/fall time is 3--5 ns, with a maximum mismatch of no more than 0.5 ns.
\end{enumerate}

The capacity calculation uses these four requirement groups because they directly constrain the time-domain NAIM output waveform. Return loss, transformer droop, duty-cycle distortion (DCD), and jitter are additional 100BASE-TX transmitter tests that characterize properties beyond the four waveform constraints used here. Observation resolution then determines the minimum parameter change counted as distinguishable.

The symbols used for the directional step size, feasible range, and state count are listed in Table I.

\begin{table}[!t]
\centering
\caption{Symbols Used for Fingerprint-Space Capacity}
\label{tab:capacity-symbols}
\renewcommand{\arraystretch}{1.3}
\begin{tabular}{@{}lp{0.68\columnwidth}@{}}
\toprule
Symbol & Description \\
\midrule
\(\mathbf{q}\) & Joint coordinate, \(\mathbf{q}=\bigl[z_{A_{\mathrm{odd}}},\,z_{A_{\mathrm{even}}},\,\mathbf{z}_{\mathrm{BP,res}}^{\mathrm{T}}\bigr]^{\mathrm{T}}\) in \eqref{eq:q} \\
\(v_{i}\) & \(i\)-th right singular vector of \(\overline{J}_{q}\) in \eqref{eq:svd} \\
\(\delta_{i}(k_{\sigma})\) & Directional step size, \(\delta_{i}(k_{\sigma})=k_{\sigma}\sigma_{i}\) in \eqref{eq:step} \\
\(r_{i}\) & Feasible range along \(v_{i}\), \(r_{i}=\alpha_{i,+}+\alpha_{i,-}\) in \eqref{eq:range} \\
\(d\) & Number of retained directions \\
\(\rho\) & Normalized scan radius of \(B_{d}(\rho)\) \\
\(p_{\mathrm{joint}}(\rho)\) & Joint feasibility rate in \eqref{eq:pjoint} \\
\(N_{\mathrm{axis}}\) & State count with independent directions, \eqref{eq:nspace} \\
\(N_{\mathrm{space}}\) & Number of jointly feasible distinguishable states, \eqref{eq:nspace} \\
\bottomrule
\end{tabular}
\end{table}

\subsection{Observation Resolution and Directional Step Size}\label{c.-observation-resolution-and-directional-step-size}

Waveform feasibility alone does not determine the number of distinguishable states; the observation model also sets the minimum parameter change counted as distinguishable. We model this resolution using additive noise and quantization by an analog-to-digital converter (ADC) with bit depth \(B_{\mathrm{ADC}}\) and full-scale range \(V_{\mathrm{pp}}\). For a specified signal-to-noise ratio (SNR), the signal and noise powers satisfy

\begin{equation}
\mathrm{SNR}_{\mathrm{dB}} = 10\log_{10}\frac{P_{s}}{P_{n}},\qquad P_{n} = P_{s}\, 10^{- \mathrm{SNR}_{\mathrm{dB}}/10},
\label{eq:snr}
\end{equation}

where \(P_{s}\) is the average power of the noiseless waveform and \(P_{n}\) is the average power of the added white noise. Following the uniform-quantization model [24], the ADC quantization step is

\begin{equation}
\Delta_{\mathrm{ADC}} = \frac{V_{\mathrm{pp}}}{2^{B_{\mathrm{ADC}}} - 1}.
\label{eq:delta-adc}
\end{equation}

The waveform requirements determine whether the noiseless NAIM output \(y_{0}\) is admissible, whereas additive and quantization noise determine the directional step size \(\delta_{i}\) used to count distinguishable states. The corresponding \(B_{\mathrm{ADC}}\), \(V_{\mathrm{pp}}\) and SNR settings are given in Section V.

The observation model treats additive noise and quantization error as independent samplewise perturbations with a common variance. Under this model, waveform changes can be compared directly after normalization by the scalar observation-noise standard deviation.

Let \(J_{q}\) be the Jacobian of the noiseless waveform with respect to \(\mathbf{q}\), and let \(\sigma_{n}^{2} = P_{n} + \Delta_{\mathrm{ADC}}^{2}/12\) denote the observation-noise variance. The noise-normalized Jacobian is obtained by singular value decomposition (SVD) as

\begin{equation}
\overline{J}_{q} = \frac{J_{q}}{\sigma_{n}} = \mathbf{U}\boldsymbol{\Sigma}\mathbf{V}^{\mathrm{T}},
\label{eq:svd}
\end{equation}

where \(\mathbf{U}\) and \(\mathbf{V}\) contain the left and right singular vectors, respectively, and the diagonal entries of \(\boldsymbol{\Sigma}\) are arranged in descending order.

The columns of \(\mathbf{V}\) define orthogonal directions \(v_{i}\) in the \(\mathbf{q}\)-coordinate space and are ordered by decreasing singular value. For a unit perturbation along \(v_{i}\), the corresponding singular value equals the norm of the resulting waveform change after normalization by the observation-noise standard deviation.

For each direction, \(s_{i} = J_{q} v_{i}\) denotes the corresponding waveform perturbation. Projecting the additive and quantization noise realizations onto \(s_{i}\) yields the directional noise standard deviation used to determine the step size in \eqref{eq:step}.

\begin{equation}
\sigma_{i} = \operatorname{std}\left( \frac{\langle n,s_{i}\rangle}{\langle s_{i},s_{i}\rangle} \right),\qquad\delta_{i}(k_{\sigma}) = k_{\sigma}\sigma_{i}.
\label{eq:step}
\end{equation}

where \(n\) is an observation-noise realization, \(\delta_{i}\) is the directional step size, \(\sigma_{i}\) is the corresponding directional noise standard deviation, and \(k_{\sigma}\) is a dimensionless resolution multiplier. The numerical setting of \(k_{\sigma}\) is given in Section V.

With a fixed timing offset and sampling phase, for a transition symbol, the minimum distance to the thresholds \(\pm 0.5\) V is computed as \(m_{\min}\). The noise standard deviation under the same observation model is denoted \(\sigma_{BER}\). The Gaussian tail function Q(·) gives the one-sided threshold [25], while the target BER follows the adopted BASE-T BER objective [26]. Define the decision-margin ratio

\begin{equation}
\eta_{\mathrm{dec}} = \frac{m_{\min}}{\sigma_{\mathrm{BER}}} \geq Q^{- 1}(10^{- 9}).
\label{eq:margin}
\end{equation}

Equation~\eqref{eq:margin} provides an auxiliary check of whether a feasible waveform retains the target decision margin under the timing and noise assumptions specified in Section V.

\subsection{Jointly Feasible States and Fingerprint-Space Capacity}\label{d.-jointly-feasible-states-and-fingerprint-space-capacity}

Along each direction \(v_{i}\), the coordinate \(\mathbf{q}\) is displaced in both signs and the resulting NAIM waveform is checked against the four time-domain requirements. The largest feasible amplitudes are denoted \(\alpha_{i, +}\) and \(\alpha_{i, -}\). The one-dimensional range is

\begin{equation}
r_{i} = \alpha_{i, +} + \alpha_{i, -}.
\label{eq:range}
\end{equation}

A combination that stays inside each one-dimensional range can still violate the waveform requirements when several directions vary together, because the retained directions act through the same output waveform.

Fig. 3 shows a two-dimensional cross-section of this coupling, with the other coordinates fixed.

\begin{figure}[!t]
\centering
\includegraphics[width=\columnwidth]{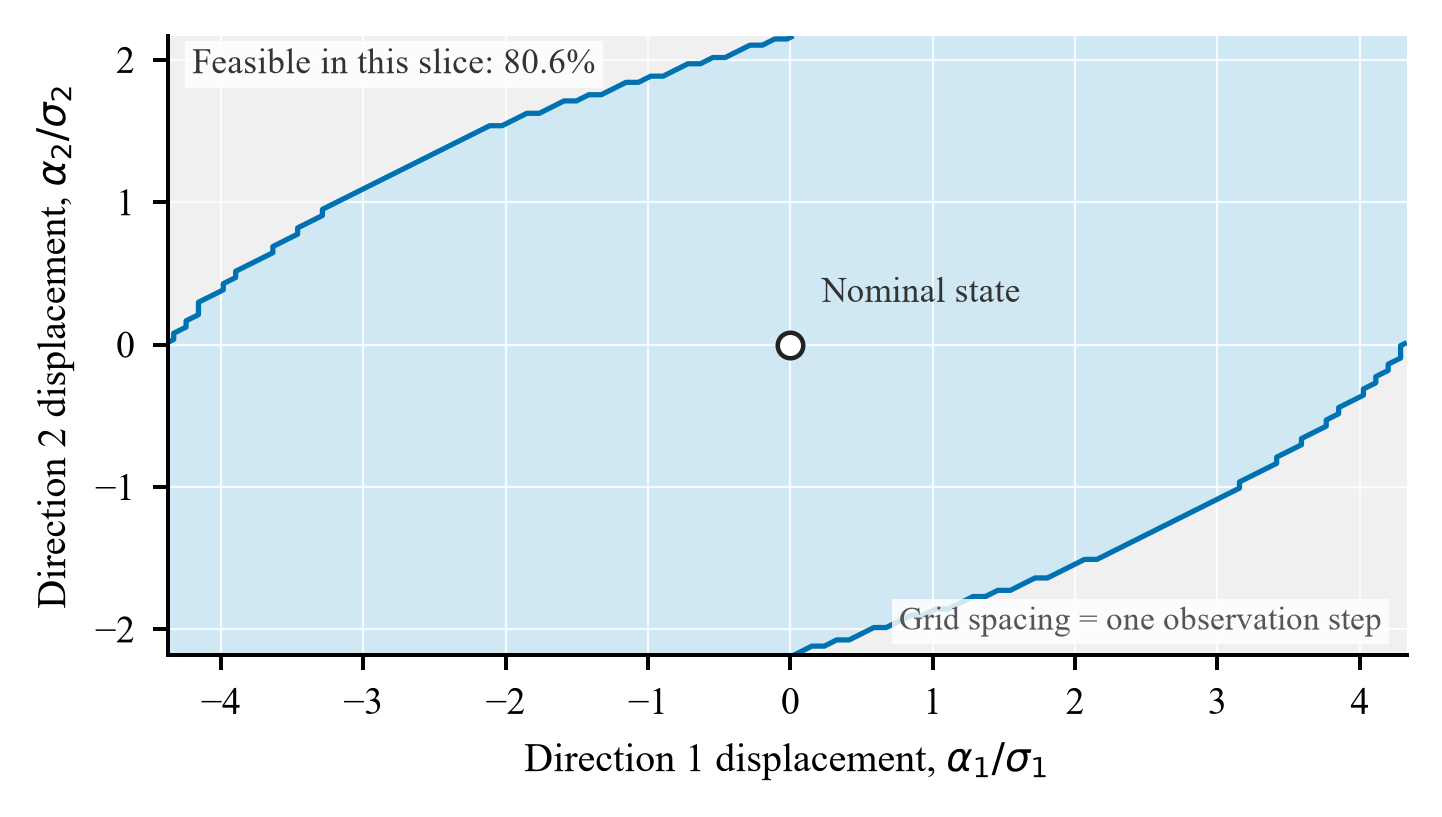}
\caption{Two-dimensional illustrative cross-section of the joint feasible region.}
\end{figure}

In Fig. 3, the boundary is oblique: the allowable range along one direction depends on the other. The gray region contains combinations that violate the time-domain requirements. The main boundary in this cross-section corresponds to waveforms that remain more than 1\% away from the steady-state value 8 ns after the transition begins. The grid spacing is one step \(\delta_{i}\). The one-dimensional ranges therefore cannot be multiplied to obtain the number of jointly feasible states. Fingerprint-space capacity uses these ranges together with the fraction of combinations that remain feasible.

A direction enters the capacity calculation when both waveform boundaries are finite and its feasible range spans more than one resolution step at the reference resolution, i.e., \(r_{i}/\delta_{i}(1)>1\). The SVD-ordered indices of these directions are denoted as \(\mathcal{A}\), and \(\mathcal{A}_{d}\) consists of the first \(d\) entries of \(\mathcal{A}\). The continuous equivalent state count along direction \(i\) is

\begin{equation}
N_{i}(k_{\sigma}) = r_{i}/\delta_{i}(k_{\sigma}).
\label{eq:ni}
\end{equation}

If the retained directions were independent, these counts would multiply to \(N_{\mathrm{axis}}\). The waveform requirements, however, couple these directions through the common output waveform. Forming the Cartesian product of the retained directional intervals and scaling it by the normalized scan radius \(\rho\) yields \(B_{d}(\rho)\). \(F_{d}(\rho)\) is the feasible subset of \(B_{d}(\rho)\). The joint feasibility rate is then

\begin{equation}
p_{\mathrm{joint}}(\rho) = \frac{\operatorname{Vol}\bigl( F_{d}(\rho) \bigr)}{\operatorname{Vol}\bigl( B_{d}(\rho) \bigr)}.
\label{eq:pjoint}
\end{equation}

The joint feasibility rate generally has no closed-form expression and is therefore estimated by randomized quasi-Monte Carlo (RQMC) sampling with scrambled Sobol sequences [27]. Each scramble maps Sobol points from the unit cube onto \(B_{d}(\rho)\) and tests the generated NAIM waveforms against the waveform requirements. For the \(r\)-th independent scramble, the feasible-point fraction is

\begin{equation}
\hat{p}_{\mathrm{joint},r} = \frac{N_{\mathrm{feasible},r}}{N_{\mathrm{Sobol},r}}
\label{eq:phat-r}
\end{equation}

where \(N_{\mathrm{Sobol},r}\) is the number of Sobol samples in the \(r\)-th scramble and \(N_{\mathrm{feasible},r}\) is the corresponding number of jointly feasible samples. The average across the \(R\) independent scrambles is

\begin{equation}
\hat{p}_{\mathrm{joint}}(\rho) = \frac{1}{R}\ \sum_{r = 1}^{R}\hat{p}_{\mathrm{joint},r}.
\label{eq:phat}
\end{equation}

\begin{equation}
\begin{aligned}
N_{\mathrm{axis}}(\rho,k_{\sigma})
&=\prod_{i\in\mathcal{A}_{d}}
\max\Bigl(1,\frac{\rho r_{i}}{\delta_{i}(k_{\sigma})}\Bigr),\\
N_{\mathrm{space}}(\rho,k_{\sigma})
&=\hat{p}_{\mathrm{joint}}(\rho)\,N_{\mathrm{axis}}(\rho,k_{\sigma}).
\end{aligned}
\label{eq:nspace}
\end{equation}

In \eqref{eq:nspace}, \(\max(1, \cdot )\) prevents any retained direction from contributing fewer than one state. \(N_{\mathrm{axis}}\) is the state count obtained when the retained directions are treated as independent. The reported \(N_{\mathrm{space}}\) is obtained by substituting the RQMC estimate \(\hat{p}_{\mathrm{joint}}(\rho)\) into \eqref{eq:nspace}.

Let \(N_{\mathrm{space,NL}}\) denote the count associated with the two steady-state coordinates alone. Including the residual impulse-response coordinates increases this count by a factor of approximately 78. Together, the waveform requirements, observation resolution, and joint feasibility rate determine the fingerprint-space capacity. Equation~\eqref{eq:margin} is used to check the decision margin of the feasible states.

\FloatBarrier
\section{Experimental Setup and Evaluation}\label{v.-experimental-setup-and-evaluation}

The experiments use waveforms collected from the 48 NICs to evaluate NAIM reconstruction, to estimate empirical capacity under the two cable conditions, and to perform closed-set identification at 5 m. The numerical settings of the fingerprint-space calculation are given in this section.

\subsection{Experimental Platform and Data Acquisition}\label{a.-experimental-platform-and-data-acquisition}

The experiments use 48 physical network interface cards (NICs) from three models---AX88772, CH9151, and RTL8152B---with 16 devices per model. Measurements span a four-day campaign. Waveforms for the 0.5-m cable condition were collected on Days 1, 2, and 4, and those for the 5-m condition on Days 2, 3, and 4. The third 0.5-m acquisition (Day 4) used a different physical cable arrangement from the first two acquisitions. For every device, cable condition, and measurement day, two consecutive oscilloscope captures, A and B, were recorded.

The oscilloscope records the two single-ended taps and forms the observed differential waveform as

\begin{equation}
y_{\mathrm{obs}}[k] = v_{\mathrm{CH1}}[k] - v_{\mathrm{CH2}}[k].
\label{eq:obs-diff}
\end{equation}

Fig. 4 shows the measurement path used to acquire the differential waveform at the observation point.

\begin{figure}[!t]
\centering
\includegraphics[width=\columnwidth]{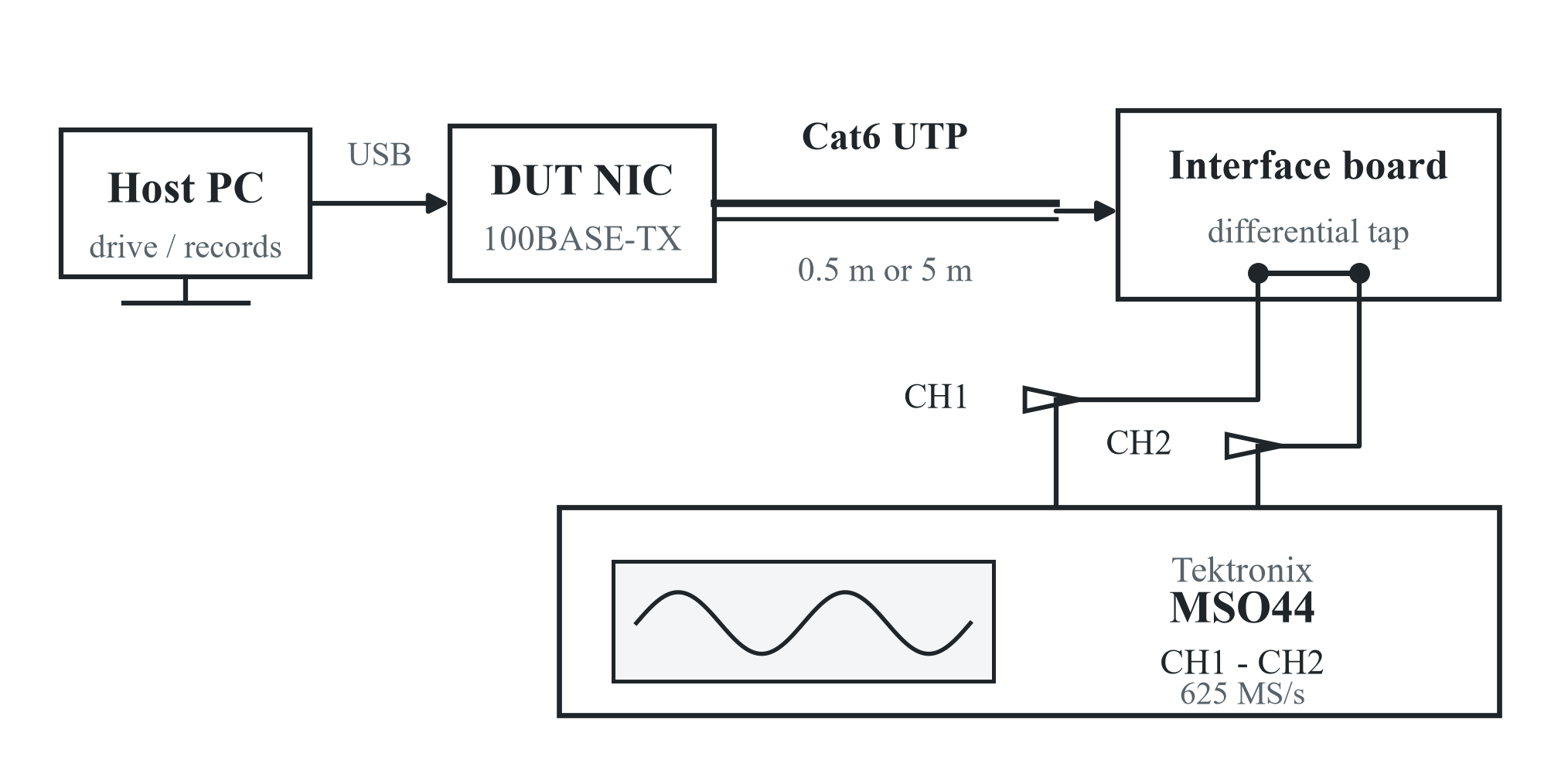}
\caption{Experimental measurement setup.}
\end{figure}

A host drives the USB--100BASE-TX NIC under test. The signal passes through the 0.5-m or 5-m Cat6 UTP cable to the interface board and is acquired by two probes connected to CH1 and CH2 of a Tektronix MSO44. The differential waveform is formed as CH1\ensuremath{-}CH2 and read back to the same host. This observation includes the joint response of the NIC, the cable, the terminal load, and the measurement chain.

The differential waveforms recorded in this observation are the data used for model reconstruction, empirical-capacity estimation, and closed-set identification. Model reconstruction and empirical-capacity estimation use all three measurement days available for each cable condition. Closed-set identification uses the 5-m measurements in a leave-one-day-out evaluation. Captures A and B provide repeated observations within a day, while the held-out day evaluates variation across days.

Table II summarizes the experimental and numerical settings.

\begin{table}[!t]
\centering
\caption{Experimental and Numerical Settings}
\label{tab:settings}
\renewcommand{\arraystretch}{1.3}
\begin{tabular}{@{}>{\raggedright\arraybackslash}p{0.28\columnwidth}p{0.66\columnwidth}@{}}
\toprule
Item & Setting \\
\midrule
Device set & 48 NICs; 16 each of AX88772, CH9151, and RTL8152B \\
Cable schedule & 0.5 m Cat6: Days 1, 2, 4 (Day 4 rewired); 5 m Cat6: Days 2, 3, 4; two captures per device and day \\
Acquisition & Tektronix MSO44; CH1\ensuremath{-}CH2; 625 MS/s; 500,000 samples; 30 waveforms per capture \\
Response model & 81-tap ridge FIR; analysis band 1--200 MHz; $M=44$ \\
Basis dictionary & Centers 12.5--200 MHz at 12.5-MHz spacing; \mbox{$Q\in\{1.5,2.5,4,7,12,20\}$}; two quadrature phases; correlation threshold 0.995 \\
Observation model & 12-bit ADC; $V_{\mathrm{pp}}=2.5$ V; SNR $=23$ dB; 1,000 noise realizations \\
Capacity estimate & $d=16$, $\rho=1$, $k_{\sigma}=1$; 8 scrambled Sobol sets $\times$ 32,768 points; $A_0=1.302\times10^{4}$ counts, $g=7.678\times10^{-5}$ V/count \\
Decision-margin check & BER $=10^{-9}$; offset 8 UI; sampling phase 0.4 UI \\
Empirical capacity & 2,000 device bootstraps; OAS for pooled within-device covariance \\
Identification & 5 m; leave-one-day-out; OAS--Mahalanobis nearest centroid \\
\bottomrule
\end{tabular}
\end{table}

\subsection{Impulse-Response Extraction and Model Configuration}\label{b.-impulse-response-extraction-and-model-configuration}

Each NIC model uses a single timing offset, which remains fixed during feature extraction. Let \(\mathbf{X}\) denote the MLT-3 convolution matrix. An 81-tap FIR response is estimated from each recorded waveform by ridge regression with \(\lambda_{\mathrm{ridge}}=10^{-6}\max[\operatorname{tr}(\mathbf{X}^{\mathsf T}\mathbf{X})/81,1]\), then mean-removed and normalized to unit \(\ell_{2}\) norm. The selected analysis band remains below the narrower oscilloscope-channel bandwidth while excluding the DC component and slow drift.

The basis waveforms are ranked by the magnitudes of their ridge-projection coefficients on a band-limited reference response and the grand-mean estimated response. Candidates that correlate strongly with an already selected waveform are discarded; in this study, the correlation threshold is set to 0.995.

The impulse-response expansion uses \(M = 44\) basis terms. In this study, \(M\) is selected by a one-standard-error rule on the reconstruction normalized mean-square error (NMSE) over the 1--200 MHz analysis band. Let \(v\) denote the estimated FIR response of a capture, \(h_{M}\) the reconstruction with \(M\) basis terms, \(\mathcal{F}\left\{ \cdot \right\}\) the Fourier transform, and \(\mathcal{B}\) the set of frequency bins within 1--200 MHz. The NMSE is

\begin{equation}
\mathrm{NMSE}_{\mathcal{B}}(M) = \frac{\sum_{f \in \mathcal{B}}\bigl| \mathcal{F}\{ v - h_{M} \}(f) \bigr|^{2}}{\sum_{f \in \mathcal{B}}\bigl| \mathcal{F}\{ v \}(f) \bigr|^{2}}.
\label{eq:nmse}
\end{equation}

The impulse-response expansion \eqref{eq:impulse} is written on this analysis band: it contains the selected basis waveforms and the shared baseline \(h_{0}\). The shared baseline \(h_{0}\) is a unit-gain Gaussian-like kernel with a 4-ns 10\%--90\% rise time. Each estimated FIR response is then projected onto this fixed basis over 1--200 MHz. The corresponding numerical settings are summarized in Table II.

\subsection{Fingerprint-Space and Gaussian-Equivalent Empirical Capacity}\label{c.-fingerprint-space-and-gaussian-equivalent-empirical-capacity}

Fingerprint-space capacity quantifies the distinguishable waveform states permitted by NAIM under the selected waveform requirements and observation resolution. Empirical capacity quantifies inter-device spread in the measured device set relative to repeated-measurement variation.

The fingerprint-space calculation uses the coordinate system in \eqref{eq:q} to vary the NAIM parameters, and the resulting waveforms are evaluated against the waveform requirements. The two steady-state parameters are converted to volts by \eqref{eq:theta-volt}, which requires a reference amplitude supplied by a reference model. The reference model is constructed from the first-day 0.5-m measurements by averaging the repeated captures within each device and then across the 48 devices. The resulting reference amplitude \(A_{0}\) is the average of the positive and negative step amplitudes of this reference waveform, equal to \(1.302\times10^{4}\) counts, giving \(g = V_{0}/A_{0} = 7.678 \times 10^{-5}\,\mathrm{V/count}\) for \(V_{0} = 1\,\mathrm{V}\). This conversion is used in \eqref{eq:theta-volt}. Observation resolution is evaluated from additive and quantization noise according to Section IV-C, while the decision-margin check follows \eqref{eq:margin}. The observation, scan, and RQMC settings are summarized in Table II.

The fingerprint-space estimate uses the first \(d=16\) active directions in the SVD ordering. Confidence intervals for the joint feasibility estimate are computed across the independently scrambled Sobol sets.

Empirical capacity is computed from the coordinate \eqref{eq:q} on the measured waveforms. Empirical capacity is estimated from one capture-level fingerprint vector \(\mathbf{f}_{\mathrm{emp}}\) per capture. Let \(\mathbf{q}_{r}\) denote the coordinate vector in \eqref{eq:q} for the \(r\)-th waveform of a capture. The capture-level fingerprint is

\begin{equation}
\mathbf{f}_{\mathrm{emp}} = \frac{1}{30}\ \sum_{r = 1}^{30}\mathbf{q}_{r}.
\label{eq:femp}
\end{equation}

The mappings in \eqref{eq:bp-reg}--\eqref{eq:q} are fitted on the 0.5-m capture-level fingerprints and applied unchanged to the 5-m fingerprints. For each cable condition in the conditioned coordinate space, \(\boldsymbol{\Sigma}_{\mathrm{dev}}\) describes the covariance of the device centroids and \(\mathbf{W}\) describes pooled variation within devices, including variation across the three measurement days, following the classical between-class and within-class statistical formulation [28]. Oracle approximating shrinkage (OAS) is applied to \(\mathbf{W}\) [29]. Under a multivariate Gaussian second-order approximation, the Gaussian-equivalent empirical capacity (hereafter empirical capacity) is the distinguishable-device count

\begin{equation}
N_{\mathrm{eq}}
=\Bigl(\prod_{i}(1+\lambda_{i})\Bigr)^{1/2},
\label{eq:neq-prod}
\end{equation}

where \(\lambda_{i}\) are the generalized eigenvalues of the covariance pair \((\boldsymbol{\Sigma}_{\mathrm{dev}}, \mathbf{W})\), and the product runs over all such eigenvalues. Equivalently,

\begin{equation}
N_{\mathrm{eq}}
=\bigl(\det(\mathbf{I}+\mathbf{W}^{-1/2}\boldsymbol{\Sigma}_{\mathrm{dev}}\mathbf{W}^{-1/2})\bigr)^{1/2}.
\label{eq:neq-det}
\end{equation}

A larger \(N_{\mathrm{eq}}\) indicates that inter-device differences occupy a broader space relative to the repeated-measurement variation within each device. Uncertainty is assessed by resampling whole devices 2,000 times. The same resampled device indices are used for both cable conditions, yielding percentile intervals for each capacity and a paired interval for the 5 m / 0.5 m empirical-capacity ratio [30]. The bootstrap resamples devices while retaining the measurement days and cable arrangements observed in the dataset; its paired interval therefore quantifies uncertainty in that ratio for this acquisition design.

\subsection{Cross-Day Closed-Set Identification}\label{d.-cross-day-closed-set-identification}

Cross-day closed-set identification evaluates whether the coordinates in \eqref{eq:q} preserve device-level separability across days. An OAS--Mahalanobis nearest-centroid classifier is used to account for unequal coordinate variances and correlations [29], [31].

Closed-set identification uses a leave-one-day-out protocol at 5 m: two days are used for registration and the remaining day for testing, with each day serving as the test day once. Feature extraction applies the fixed NIC-model timing offset described in Section V-B.

Each fold uses the two registration days to estimate device centroids and within-class covariances. For device \(j\), the training waveforms are averaged to obtain the centroid \(\mathbf{c}_{j}\). The global covariance \(\mathbf{W}_{\mathrm{OAS}}\) is estimated from the deviations of all training waveforms about these centroids and is then OAS-shrunk [29]. The squared Mahalanobis distance, and the resulting global device prediction, are

\begin{equation}
\begin{aligned}
D_{j}(\mathbf{z})&=(\mathbf{z}-\mathbf{c}_{j})^{\mathrm{T}}\mathbf{W}_{\mathrm{OAS}}^{-1}(\mathbf{z}-\mathbf{c}_{j}),\\
\hat{j}_{\mathrm{g}}&=\arg\min_{j\in\mathcal{K}}\bigl[D_{j}(\mathbf{z}_{A})+D_{j}(\mathbf{z}_{B})\bigr],
\qquad \hat{\ell}=\mathcal{L}(\hat{j}_{\mathrm{g}}),
\end{aligned}
\label{eq:dist-glob}
\end{equation}

where \(\mathbf{z}_{A}\) and \(\mathbf{z}_{B}\) are the test-day capture averages of the 30 waveforms in captures A and B, respectively, and \(\mathcal{L}(j)\) is the NIC model of registered device \(j\). The true model of a test device is not used for routing.

Within the predicted model, \(\mathbf{W}_{\hat{\ell},\mathrm{OAS}}\) is estimated in the same way from the training waveforms of that model's 16 registered devices. The local distance and the final decision are

\begin{equation}
\begin{aligned}
D_{j,\hat{\ell}}(\mathbf{z})&=(\mathbf{z}-\mathbf{c}_{j})^{\mathrm{T}}\mathbf{W}_{\hat{\ell},\mathrm{OAS}}^{-1}(\mathbf{z}-\mathbf{c}_{j}),\\
\hat{j}&=\arg\min_{j\in\mathcal{K}_{\hat{\ell}}}\bigl[D_{j,\hat{\ell}}(\mathbf{z}_{A})+D_{j,\hat{\ell}}(\mathbf{z}_{B})\bigr].
\end{aligned}
\label{eq:dist-loc}
\end{equation}

\FloatBarrier
\section{Experimental Results}\label{vi.-experimental-results}

\subsection{Reconstruction Results of the Nonlinear and Impulse-Response Model (NAIM)}\label{a.-reconstruction-results-of-the-nonlinear-and-impulse-response-model-naim}

With \(M=44\), NAIM reconstructs the estimated 1--200 MHz response under both cable conditions. Table III summarizes the reconstruction error and the in-band energy of the estimated FIR response.

\begin{table}[!t]
\centering
\caption{Model Reconstruction Performance Under the Two Cable Conditions}
\renewcommand{\arraystretch}{1.3}
\begin{tabular}{@{}lccc@{}}
\toprule
Cable & $N$ & NMSE & In-band energy \\
\midrule
0.5 m & 288 & 0.023 [0.001, 0.046] & 72.2\% \\
5 m & 288 & 0.033 [0.002, 0.047] & 67.6\% \\
\bottomrule
\end{tabular}

\vspace{2pt}
{\raggedright\scriptsize Note: $N$ is the number of captures per cable condition ($48\times 2$ (A/B)$\times 3=288$); NMSE is the median [interquartile range]; in-band energy is the median.\par}
\end{table}

The median NMSEs for 0.5 m and 5 m are 0.023 and 0.033, respectively, with the upper quartile not exceeding 0.047 in either case; the median in-band energies are 72.2\% and 67.6\%, respectively. Fig. 5 shows the time-domain and frequency-domain reconstruction results for representative captures.

\begin{figure*}[!t]
\centering
\includegraphics[width=\textwidth]{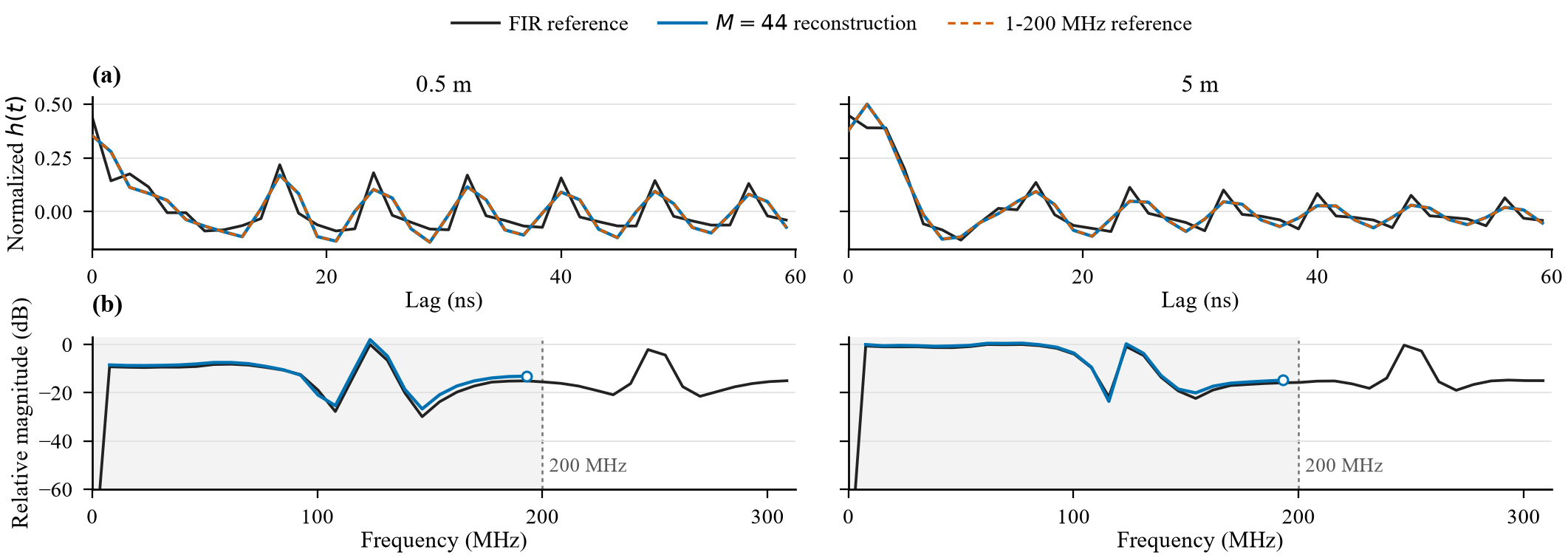}
\caption{Time-domain and frequency-domain reconstruction results. (a) Band-limited estimated FIR response and model reconstruction for representative captures under the 0.5-m and 5-m cable conditions. (b) Corresponding spectra.}
\end{figure*}

As shown in Fig. 5(a), the reconstruction follows the band-limited estimated FIR response in the initial peak, subsequent peak locations, and decay. Fig. 5(b) shows agreement in the main spectral peaks and notches over 1--200 MHz. Together with Table III, these results indicate that the 44 impulse-response parameters capture the main temporal and spectral structure of the estimated in-band response under both cable conditions.

\subsection{Fingerprint-Space Capacity and Industrial-Scale Context}\label{b.-fingerprint-space-capacity-and-industrial-scale-context}

The first 16 SVD directions account for 99.991\% of the total squared singular-value sum. Within the retained 16-direction subspace, the model-based estimate is approximately 2.96 \ensuremath{\times} $10^{10}$ distinguishable states at \(\rho = 1\) and \(\ensuremath{k_{\sigma}} = 1\):

\begin{equation}
\begin{aligned}
\hat{p}_{\mathrm{joint}}&=\frac{417}{8\times 32768}\approx 1.591\times 10^{-3},\\
N_{\mathrm{axis}}&\approx 1.86\times 10^{13},\\
N_{\mathrm{space}}&=N_{\mathrm{axis}}\,\hat{p}_{\mathrm{joint}}\approx 2.96\times 10^{10}.
\end{aligned}
\label{eq:nspace-num}
\end{equation}

Ignoring direction coupling yields approximately \(1.86\times10^{13}\) distinguishable states, whereas enforcing joint waveform feasibility yields approximately \(2.96\times10^{10}\) distinguishable states. This reduction by a factor of approximately 628 quantifies the effect of directional coupling: parameter combinations that are admissible along individual directions can violate the waveform requirements when varied jointly. The resulting jointly feasible states also satisfy the decision-margin criterion in \eqref{eq:margin} under the settings in Table II.

Fig. 6 shows how the fingerprint-space capacity varies with the normalized scan radius \ensuremath{\rho} within the retained 16-direction subspace.

\begin{figure}[!t]
\centering
\includegraphics[width=0.95\columnwidth]{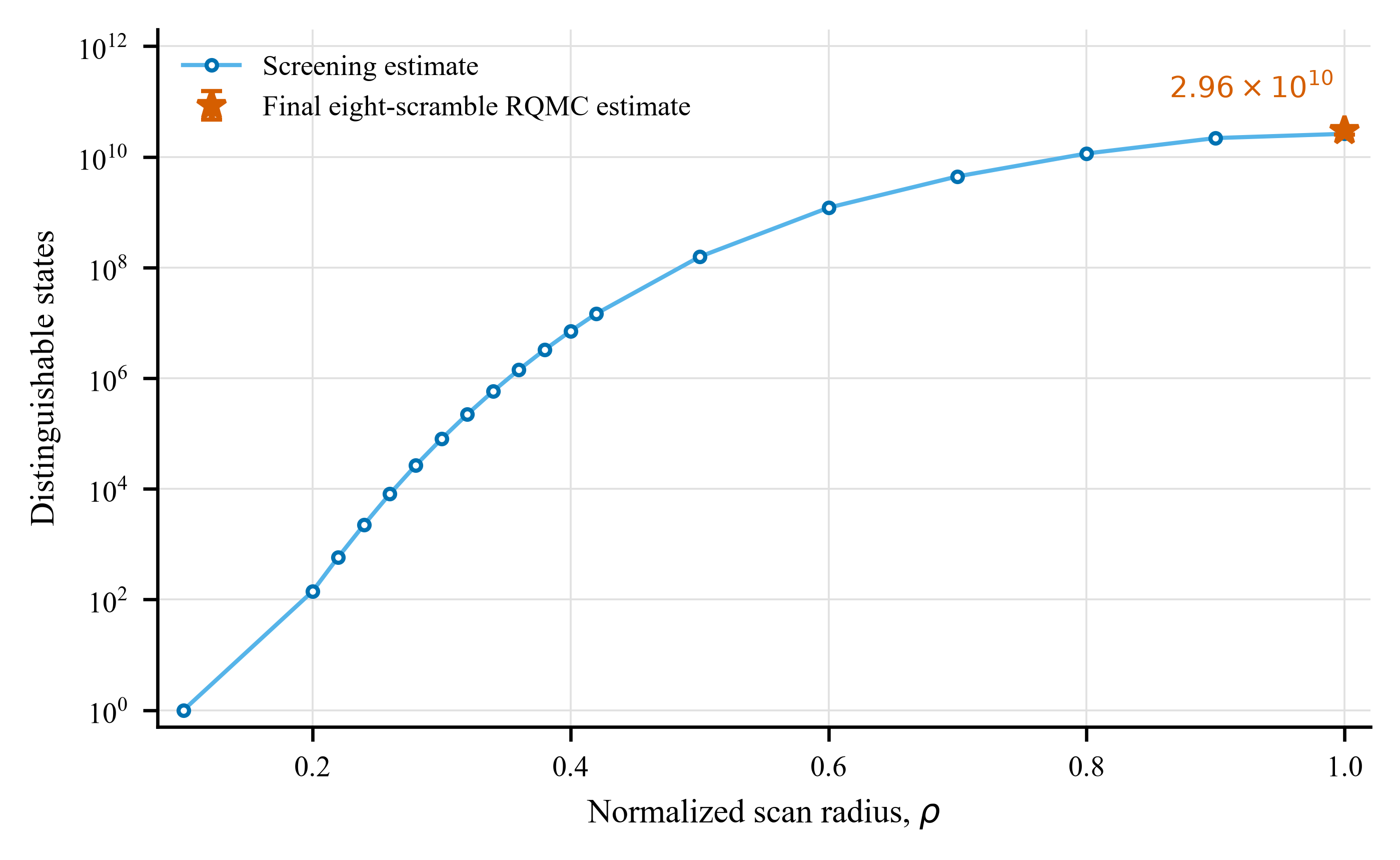}
\caption{Fingerprint-space capacity versus normalized scan radius $\rho$ in the retained 16-direction subspace. The curve shows the capacity estimates at different values of $\rho$, and the star denotes the main result at $\rho=1$.}
\end{figure}

As shown in Fig. 6, the estimated capacity increases as \(\rho\) expands the portion of the retained parameter range included in the scan. At \(\rho = 1\) the fingerprint-space capacity is approximately \(2.96\times10^{10}\) distinguishable states.

The eight RQMC estimates yield a 95\% confidence interval of \([2.54\times10^{10},\ 3.37\times10^{10}]\) distinguishable states, preserving the \(10^{10}\)-state scale. The required separation between states has a much stronger effect: requiring two and three times the baseline separation (\(k_{\sigma}=2\) and 3) reduces the capacity to approximately \(4.51\times10^{5}\) and 687 distinguishable states, respectively.

The two steady-state level coordinates yield approximately 78 distinguishable states at \ensuremath{\rho}=1 and \ensuremath{k_{\sigma}}=1. Adding the residual impulse-response coordinates gives the full fingerprint-space capacity of approximately \(2.96\times10^{10}\) distinguishable states.

For an order-of-magnitude industrial reference, the EtherCAT Technology Group reports about 105.2 million accumulated EtherCAT nodes by 2025, while PROFIBUS \& PROFINET International reports about 89.2 million accumulated PROFINET nodes [32], [33]. These deployments are both on the order of \(10^{8}\) nodes, whereas the model-based fingerprint-space estimate is approximately \(2.96\times10^{10}\) \mbox{distinguishable states}.

\subsection{Gaussian-Equivalent Empirical Capacity and Measured Devices}\label{c.-gaussian-equivalent-empirical-capacity-and-measured-devices}

Table IV reports the fingerprint-space capacity under the stated waveform and observation settings, together with the empirical capacity estimated from inter-device spread relative to within-device variation for the 48 measured devices.

\begin{table}[!t]
\centering
\footnotesize
\caption{Fingerprint-Space and Empirical Capacity Estimates}
\renewcommand{\arraystretch}{1.3}
\setlength{\tabcolsep}{3pt}
\begin{tabular}{@{}lccc@{}}
\toprule
Quantity & Full & Level-only & 95\% interval \\
\midrule
\(N_{\mathrm{space}}\) (\(d=16\)) & \(2.96\times10^{10}\) & 78 & \([2.54,\ 3.37]\times10^{10}\) \\
Empirical capacity, 0.5 m & \(1.63\times10^{3}\) & 10.3 & \([0.971,\ 2.21]\times10^{3}\) \\
Empirical capacity, 5 m & \(2.44\times10^{5}\) & 21.2 & \([0.687,\ 3.21]\times10^{5}\) \\
Paired ratio, 5 m / 0.5 m & 101 & --- & \([44.6,\ 219]\) \\
\bottomrule
\end{tabular}

\vspace{2pt}
{\raggedright\footnotesize Note: Capacity entries are distinguishable-state counts; empirical capacities are Gaussian-equivalent. For \(N_{\mathrm{space}}\), the interval is a 95\% $t$ interval across eight RQMC scrambles; empirical quantities use 95\% device-bootstrap percentile intervals.\par}
\end{table}

Table IV compares the model-based fingerprint-space scale with the empirical separability of the measured devices. The fingerprint-space capacity of \(2.96\times10^{10}\) distinguishable states characterizes the distinguishable waveform-state scale permitted by NAIM under the selected waveform and observation settings. The empirical capacities of \(1.63\times10^{3}\) and \(2.44\times10^{5}\) distinguishable devices characterize the inter-device spread observed in the 48 measured NICs relative to repeated-measurement variation under the two acquisition conditions. The numerical gap reflects these different levels of characterization together with the finite measured device population and its measurement variation.

As shown in Table IV, the empirical capacities at 0.5 m and 5 m are approximately \(1.63\times10^{3}\) and \(2.44\times10^{5}\) distinguishable devices under the Gaussian model, respectively. The paired bootstrap yields a median 5 m--0.5 m empirical-capacity ratio of approximately 101, with a 95\% interval of \([44.6,\ 219]\).

Under the original measurement grouping, the 5-m data exhibit a higher separation-to-variation ratio than the 0.5-m data. The two datasets were acquired with different day compositions, and the third 0.5-m acquisition was taken after the cable was reconnected, using a different physical cable arrangement from the first two. The 0.5-m estimate is particularly sensitive to this third acquisition: omitting it increases the empirical capacity from approximately \(1.63\times10^{3}\) to \(2.05\times10^{5}\) distinguishable devices, whereas omitting either of the first two days changes it by a factor of no more than 1.52. These results show that different cable environments, even under the same receiver, significantly affect empirical capacity and practical identification performance.

\subsection{Closed-Set Identification and NIC-Model Comparison}\label{d.-closed-set-identification-and-nic-model-comparison}

Under the 5-m observation condition, the cross-day closed-set Top-1 accuracy reaches 90.97\%, with fold-wise accuracies of 89.58\%, 91.67\%, and 91.67\%.

To further compare the three NIC models, Fig. 7 visualizes the 48 device centers in the first two principal components under the 5-m observation condition.

\begin{figure}[t]
\centering
\includegraphics[width=\columnwidth]{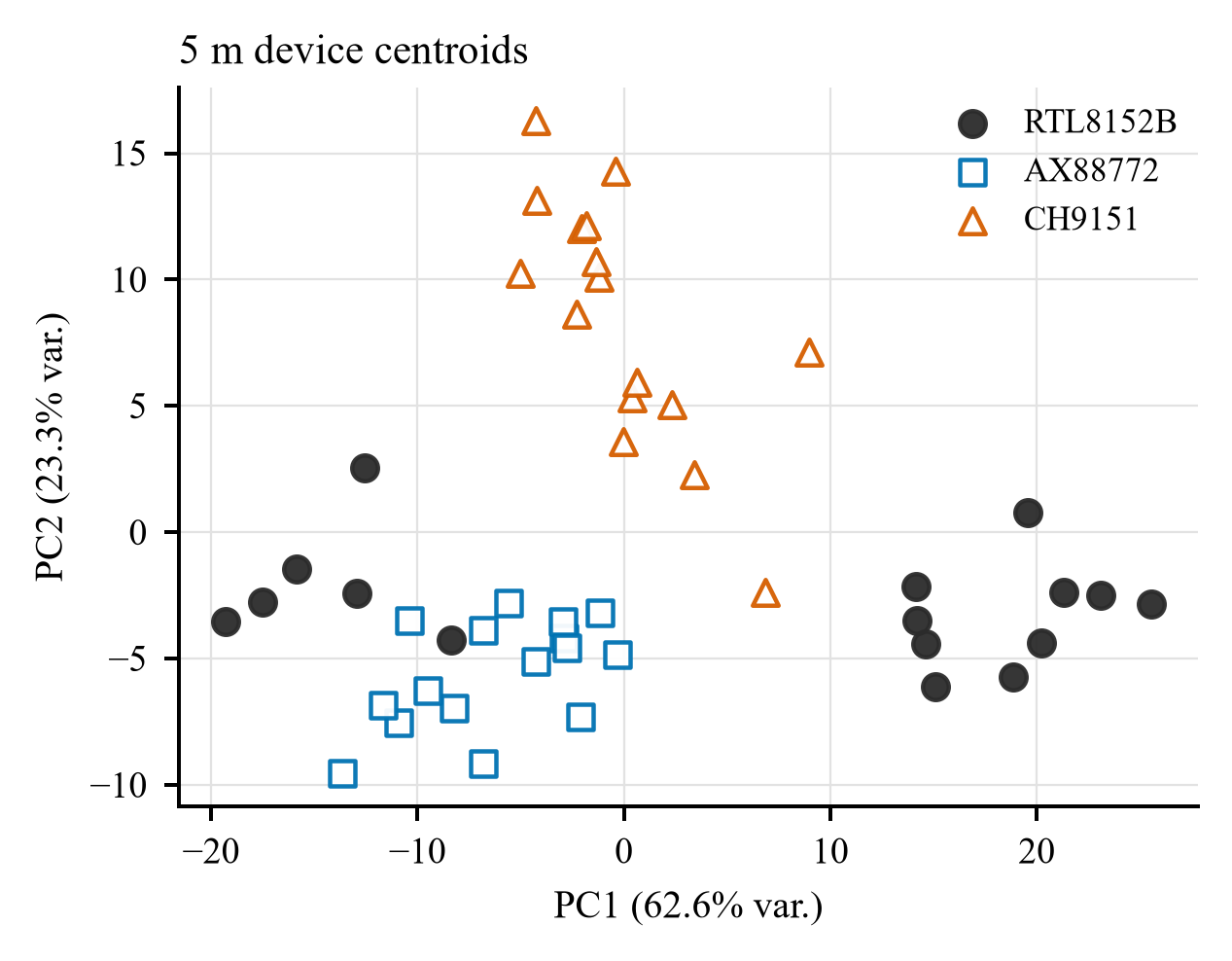}
\caption{Principal-component projection of the device centers under the 5-m condition.}
\end{figure}

As shown in Fig. 7, the three NIC models largely occupy different regions. Two sub-clusters appear within RTL8152B. The first two principal components explain about 85.9\% of the variance. This projection illustrates the distribution of device centers across the three NIC models. The quantitative comparison of empirical capacity and closed-set Top-1 accuracy is reported in Table V.

\begin{table}[t]
\centering
\caption{Empirical Capacity and Cross-Day Closed-Set Identification Under the 5-m Condition}
\renewcommand{\arraystretch}{1.3}
\begin{tabular}{@{}lccc@{}}
\toprule
Group & $N$ & Empirical capacity & Top-1 (\%) \\
\midrule
All devices & 48 & \(2.44\times10^{5}\) & 90.97\% \\
RTL8152B & 16 & \(3.98\times10^{3}\) & 97.92\% \\
AX88772 & 16 & 813 & 95.83\% \\
CH9151 & 16 & 20 & 79.17\% \\
\bottomrule
\end{tabular}

\vspace{2pt}
{\raggedright\scriptsize Note: The all-devices row summarizes the full 48-device set; the NIC-model comparison refers to the three model-specific rows.\par}
\end{table}

RTL8152B, AX88772, and CH9151 exhibit identical rankings in empirical capacity and cross-day Top-1 accuracy. This agreement establishes a direct experimental link between empirical capacity and identification performance. It demonstrates that empirical capacity captures the device separability preserved across days and provides a quantitative measure of the recognition capability of each NIC population.

\section{Conclusion}\label{vii.-conclusion}

This paper investigates the fingerprint-space capacity of 100BASE-TX physical-layer fingerprints for IIoT authentication. By analyzing their generation, a nonlinear and impulse-response model (NAIM) is developed. Based on this model, the fingerprint-space capacity of 100BASE-TX terminals is derived under the transmitter waveform requirements, observation resolution, and target BER.

The fingerprint space contains approximately \(2.96\times10^{10}\) distinguishable states, which is significantly larger than reported industrial Ethernet inventories of about \(10^{8}\) nodes. This result provides a basis for using physical-layer fingerprints for IIoT access authentication. Experiments on 48 NICs under the 5-m cable condition show that empirical capacity and closed-set identification consistently rank the three NIC models, indicating that the proposed capacity analysis provides a pre-deployment assessment for physical-layer fingerprinting and can guide the selection of NIC models.

\begin{samepage}
The same analysis applies to the fingerprint-space capacity of other industrial links, such as PLC, RS-485, and CAN, after the corresponding constraints are restated. Future work will examine temperature variation, longer-term drift, and cable-length variation, and will extend the analysis to larger device populations.
\end{samepage}

\section*{Conflict of Interest}
The authors declare that they have no conflicts of interest relevant to this work.

% references handled below\label{references}

\end{document}